# CHAOS SYNCHRONIZATION BETWEEN THE JOSEPHSON JUNCTIONS GOVERNED BY THE CENTRAL JUNCTION UNDER THE EFFECT OF PARAMETER MISMATCHES AND NOISE

**E. M. SHAHVERDIEV[1*], L. H. HASHIMOVA[1], P. A. BAYRAMOV[1],**
**R. A. NURIEV[1] AND M. V. QOCAYEVA[1]**
[1]Institute of Physics, 33, H. Javid Avenue, Baku, AZ1143, Azerbaijan.

**AUTHORS' CONTRIBUTIONS**
This work was carried out in collaboration between the all authors. Author EMS designed the study, wrote the algorithm, methodology programming and interpreted the results. Authors LHH, PAB, RAN and MVQ gathered the initial data and helped to interpret the results. All authors read and approved the final manuscript.

**ABSTRACT**

Chaos synchronization between Josephson junctions driven by a central junction is studied under the parameter mismatches and noise. It is demonstrated that chaos synchronization quality is robust to 10-15 % parameter mismatches. It is also elucidated that for the intermediary noise intensity correlation between the synchronized junctions can be even enhanced; however for larger intensities of noise synchronization quality deteriorates. It is also established that synchronization quality between some junctions remains unchanged, despite the fact that all the driven junctions and the driver junction were subject to the same amount of noise.
These results are of certain importance for obtaining the high power system of Josephson junctions in real-life situations and are promising for practical applications in the Terahertz region.

# 1 Introduction

Chaos synchronization and chaos control are hot topics in non-linear dynamics, see e.g. [1-6] and extensive references there-in. These phenomena are of immense fundamental importance in a variety of complex physical, chemical, power, biological, economical and social systems with possible application areas in secure communications, optimization of nonlinear system performance, modeling brain activity and pattern recognition phenomena, avoiding power black-out, obtaining high power radiation sources, etc.

Recently due to the search for the new Terahertz sources investigation of chaos synchronization between Josephson junctions is also becoming ever expanding research field in the non-linear physics [7].

Being situated between microwaves and infrared light waves Terahertz (THz) radiation has a huge range of possible uses, including security scanning, remote sensing chemical signatures of explosives, non-invasive applications in medicine, ecology, secure communication, etc. [8-9]. Due to this fact recently technology for generating and manipulating Terahertz radiation is the subject of intense research. Currently available sources of Terahertz radiation, such as the gyrotron, the far infrared laser, the free electron laser, quantum cascade laser, etc.[8-9], as a rule require a lot of space, not portable, and most importantly in many cases need costly deep cooling procedure. These factors make the search for the compact, cheap, and portable Terahertz sources of paramount importance. Besides, for many remote sensing and imaging applications high power Terahertz sources are vitally important [8].

Although Josephson junctions can be a source of radiation with frequencies up to Terahertz region, the radiation from a single Josephson junction is extremely weak, usually up to nW [7]. Fortunately synchronization of Josephson junctions can help to obtain a practically viable source of the radiation power [10-12]. Recently it was shown that certain highly anisotropic high temperature cuprate superconductors [7], such as BSCCO naturally contain a stack of thousands strongly electromagnetically coupled intrinsic Josephson junctions. The Josephson junctions in high-$T_c$ BSCCO crystal may serve as a viable candidate for the achieving of powerful Terahertz radiation. It is noted that in comparison with Josephson junctions in low-$T_c$ materials high-$T_c$ crystals have a large superconducting energy gap (the energy gap of BSCCO ranges from 10 to 60 meV), which can provide a frequency range up to 5-30 THz [13]. It should also be emphasized that recently the authors of [14] have demonstrated that Josephson junctions in $YBa_2Cu_3O_{7-\delta}$ can be used as a terahertz wave detectors.

In [15] we have studied chaos synchronization between unidirectionally coupled Josephson junctions. We have established that it is possible to synchronize several tens of such junctions in such a configuration. In [16] we have investigated synchronization of Josephson junctions governed by a central junction. It has been shown that with this coupling topology it is possible to synchronize several hundreds, or even probably thousands junctions. As it follows from these results, the coupling topology is important for the number of junctions to be synchronized. As mentioned above there are thousands of naturally intrinsic Josephson junctions in some of high -$T_c$ superconducting crystals. If needed, for technological purposes such high number of junctions can be engineered artificially. From this point of view it was of enormous practical importance to study further the effect of coupling topology on the synchronization quality among the Josephson junctions.

In this connection in [17] we studied synchronization between bidirectionally coupled Josephson junctions and compared the results with the case of unidirectionally coupled systems. Furthermore we also compared the results with the case of systems coupled in a mixed configuration, e.g. coupling between the first two junctions was unidirectional, while the remaining junctions were coupled bidirectionally. We established that in terms of synchronization quality unidirectionally coupled systems perform better. These findings are also especially important in the context of Josephson junctions governed by the central junction, where the coupling between the governing central junction and the rest of junctions is unidirectional.

In real situations (natural or man made) parameters of the systems to be synchronized can be changeable due to the fluctuations, noisy environment or deliberate changes. The effect of parameter mismatches and noise on the synchronization quality between the unidirectionally coupled Josephson junctions was studied in [18].

In this paper we make the first report on the effect of parameter mismatches and noise on the chaos synchronization quality between Josephson junctions governed by a central junction. We obtain that in general, synchronization is quite robust to the parameter mismatches. As for the effect of noise on the synchronization quality we find that for some intensities noise can play even constructive role enhancing the synchronization quality. For higher noise intensities the synchronization quality deteriorates. We also find that despite the noise presence synchronization quality between some Josephson junctions remains unchanged.

Thus, the aim of this paper is to demonstrate that synchronization of the Josephson junctions driven by the master Josephson junction is robust to the parameter mismatches and influence of noise. This is very important from the application view point of the terahertz radiation sources based on the synchronization.

## 2 Josephson Junctions Governed by the Central Junction with Parameter Mismatches

Consider the following model of RC-shunted Josephson junctions driven by the central junction [16] (Fig. 1 provides a schematic presentation of the synchronization scheme) written in the dimensionless form:

$$\frac{d\phi_1}{dt} = \psi_1 \tag{1}$$

$$\frac{d\psi_1}{dt} = -\beta_1\psi_1 - \sin\phi_1 + i_{dc1} + i_{01}\cos\varphi_1 \tag{2}$$

$$\frac{d\varphi_1}{dt} = \Omega_1 \tag{3}$$

$$\frac{d\phi_2}{dt} = \psi_2 \tag{4}$$

$$\frac{d\psi_2}{dt} = -\beta_2\psi_2 - \sin\phi_2 + i_{dc2} + i_{02}\cos\varphi_2 - \alpha_{s1}(\psi_2 - \psi_1(t-\tau_1)) \tag{5}$$

$$\frac{d\varphi_2}{dt} = \Omega_2 \tag{6}$$

$$\frac{d\phi_3}{dt} = \psi_3 \tag{7}$$

$$\frac{d\psi_3}{dt} = -\beta_3\psi_3 - \sin\phi_3 + i_{dc3} + i_{03}\cos\varphi_3 - \alpha_{s2}(\psi_3 - \psi_1(t-\tau_2)) \tag{8}$$

$$\frac{d\varphi_3}{dt} = \Omega_3 \tag{9}$$

$$\frac{d\phi_4}{dt} = \psi_4 \tag{10}$$

$$\frac{d\psi_4}{dt} = -\beta_4\psi_4 - \sin\phi_4 + i_{dc4} + i_{04}\cos\varphi_4 - \alpha_{s3}(\psi_4 - \psi_1(t-\tau_3)) \tag{11}$$

$$\frac{d\varphi_4}{dt} = \Omega_4 \tag{12}$$

Where $\phi_{1,2,3,4}$ are the phase differences of the superconducting order parameter across the junctions 1, 2, 3 and 4, respectively; $\psi_{1,2,3,4}$ describe the dynamics of the respective phase differences $\phi_{1,2,3,4}$.

$\beta$ is called the damping parameter $(\beta R)^2 = (h/2\pi)(2eI_c C)^{-1}$, where $I_c, R, C$ are the junctions' critical current, the junction resistance, and capacitance, respectively; $\beta$ is related to McCumber parameter $\beta_c$ by $\beta^2 \beta_c = 1$; $h$ is Planck's constant; $e$ is the electronic charge; $i_{dc1,dc2,dc3,d4}$ are the driving the junctions direct current; $i_{01,02,03,04}\cos(\Omega_{1,2,3,4}t + \theta_{1,2,3,4})$ are the driving ac (or rf) current with amplitudes $i_{01,02,03,04}$, frequencies $\Omega_{1,2,3,4}$, and phases $\theta_{1,2,3,4}$; $\tau_1$, $\tau_2$ and $\tau_3$ are the coupling delay times between the junctions 1-2, 1-3 and 1-4, respectively; couplings between the junctions 1-2, 1-3 and 1-4 are due to the currents flowing through the coupling resistors $R_{s1}$ (between the junctions 1 and 2), $R_{s2}$ between the junctions 1 and 3) and $R_{s3}$ (between the junctions 1 and 4); $\alpha_{s1,s2,s3} \sim R_{s1,s2,s3}^{-1}$ are the coupling strengths between junctions 1 - 2 junctions 1 – 3, and junctions 1-4. We note that in Eqs. (1-12) direct current and ac current amplitudes are normalized with respect to the critical currents for the relative Josephson junctions; ac current frequencies $\Omega_{1,2,3,4}$ are normalized with respect to the Josephson junction plasma frequency $\omega_{J1,J2,J3,J4} = 2eI_{c1,c2,c3,c4}((h/2\pi)C_{1,2,3,4})^{-1}$ and dimensionless time is normalized to the inverse plasma frequency. We underline that the value of phases $\theta_{1,2,3,4}$ for the ac driving forces can be incorporated into the initial conditions for Eqs.(1-12).

We note that Eqs.(1-12) are of the interdisciplinary nature and can also used for modeling of a driven nonlinear pendulum, charge density waves with a torque and sinusoidal driving field [19-20]. We also mention that there are some other ways to make the Josephson dynamics chaotic without resorting to the *ac* driving. One way is to use resistively-capacitively-inductively-shunted (RCL-shunted) Josephson junction [21]. Additionally by adding time delays in the system one can also make the Josephson junction dynamics chaotic via time- delays.

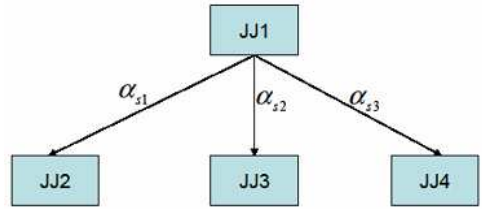

**Fig. 1. Schematic presentation of the driver Josephson junction JJ1 coupled to the driven Josephson junctions JJ2,JJ3 and JJ4.**

## 3 Numerical Simulations

First we demonstrate that three identical driven time delay coupled Josephson junctions can be synchronized by the driver (master) system. We characterize chaos synchronization quality between the Josephson junctions using the correlation coefficient $C$ [22]. This coefficient indicates the quality of synchronization: $C=1$ means perfect synchronization; lesser values of the correlation coefficient $C$ mean imperfect synchronization.

We simulate Eqs.(1-12) using the following set of parameters: $\beta_1 = \beta_2 = \beta_3 = 0.25$, $i_{dc1} = i_{dc2} = i_{dc3} = 0.3$, $i_{01} = i_{02} = i_{03} = 0.7$, $\Omega_1 = \Omega_2 = \Omega_3 = 0.6$, $\theta_1 = \theta_2 = \theta_3 = 0$, $\alpha_{s1} = \alpha_{s2} = \alpha_{s3} = 0.45$, $\tau_1 = \tau_2 = \tau_3 = 0.15$.

Fig. 2 provides dynamics of $\psi_4$. Fig. 3 demonstrates the error dynamics between the end driven junctions 2 and 4 $\psi_4 - \psi_2$, while $C_{2-4} = 1$ indicates the correlation coefficient between $\psi_2$ and $\psi_4$. The other correlation coefficients are: $C_{1-2} = 0.98$, $C_{1-3} = 0.98$, $C_{1-4} = 0.98$, $C_{2-3} = 1$, $C_{3-4} = 1$.

These simulation results underline the high quality synchronization between three driven identical Josephson junctions with different initial conditions. Note that correlation between dynamics driver (master) Josephson junction and driven (slave) Josephson junctions is slightly less (2%).

Next we present the results of the detailed study of the effect of parameter mismatches on the synchronization quality between driver (junction 1) and driven Josephson junctions (junctions 2, 3 and 4) under conditions when parameter mismatches were introduced between the driven junction 2 and 4; the parameters of the other junctions remain unchanged.

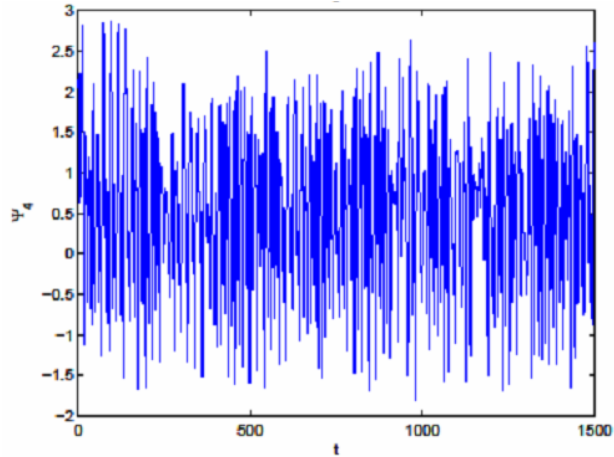

**Fig. 2.** Numerical simulation of Eqs.(1-12): dynamics of $\psi_4$. The parameters are: $\beta_1 = \beta_2 = \beta_3 = 0.25$, $i_{dc1} = i_{dc2} = i_{dc3} = 0.3$, $i_{01} = i_{02} = i_{03} = 0.7$, $\alpha_{s1} = \alpha_{s2} = \alpha_{s3} = 0.45$, $\tau_1 = \tau_2 = \tau_3 = 0.15$. Dimensionless units.

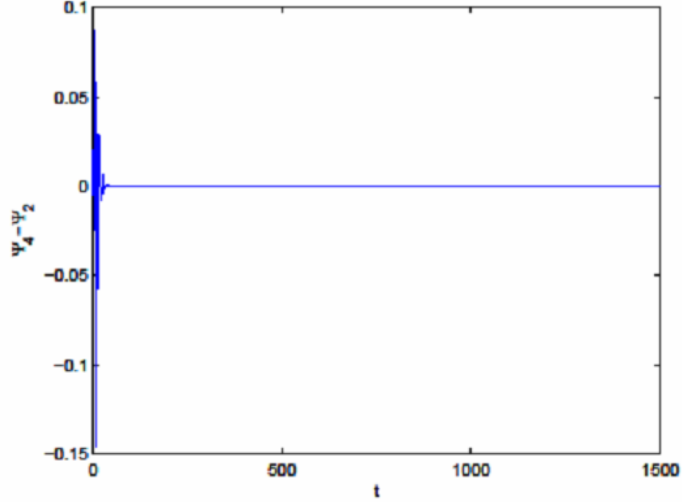

**Fig. 3.** Error dynamics of $\psi_4 - \psi_2$ : $C_{2-4}$ is the correlation coefficient between $\psi_4$ and $\psi_2$. The parameters are as for Fig. 2. Dimensionless units.

At present technologically the preparation of the identical Josephson junctions is not an impossible task. Nevertheless, mismatches (however small) between the system parameters are still a possibility due to the fluctuations, technological procedures, etc.

In the following we tabulated the results of numerical simulations mainly into the table. The reason for this is that in many cases the results for the correlation coefficients are very close which makes the figure presentations less readable and a bit less discernable.

Table 1 provides the dependence of the correlation coefficients on the ratio $\tau_3 \tau_1^{-1}$ of the coupling delay $\tau_3$ between the junctions 1 and 4 to the coupling delay time $\tau_1$ between the junctions 1 and 2. It is noted that the correlation coefficients between the driver junction and driven junctions $C_{1-2}$, $C_{1-3}$, $C_{1-4}$ are less than unity; at the same time the correlation coefficients between the driven junctions are nearly maximal. All in all 15% of parameter mismatches reduces the synchronization quality between the Josephson junctions around 1-2%.

That is the synchronization quality is quite robust to the mismatches between $\tau_3$ and $\tau_1$.

Table 2 shows the dependence of the correlation coefficients on the ratio $\alpha_{s3}\alpha_{s1}^{-1}$ of the coupling strength $\alpha_{s3}$ between the junctions 1 and 4 to the coupling strength $\alpha_{s1}$ between the junctions 1 and 2. As in the case of mismatches between $\tau_3$ and $\tau_1$, mismatches between $\alpha_{s3}$ and $\alpha_{s1}$ affects the correlation coefficients between the Josephson junctions slightly (1-2 %)

Table 3 portrays the correlation coefficients as a function of the ratio $\beta_4\beta_2^{-1}$ of the damping parameter $\beta_4$ to the damping parameter $\beta_2$. As distinct from the previous two cases in this case 15 % of parameter mismatches deteriorate the synchronization quality around 8%.

In Table 4 the dependence of the correlation coefficients on the ratio $i_{dc4}i_{dc2}^{-1}$ is shown. In this case 15 % of mismatch between the parameters decrease the correlations coefficients by 7 %.

Table 5 describes the correlation coefficients versus the ratio $i_{04}i_{02}^{-1}$. It is seen from the data presented in this table the maximal value of reduction of the correlation coefficients as a result of the parameter mismatches is about 9 %.

Presented in Table 6 data provide the correlation coefficients dependence on the ratio $\Omega_4\Omega_2^{-1}$. It seems that in this case the influence of the parameter mismatches between $\Omega_4$ and $\Omega_2$ on the correlation coefficients is quite drastic. 15% mismatch deteriorates the synchronization quality significantly, up to 52 %.

It is seen from the computer simulation results there is a slight asymmetry in the correlation coefficients data depending on the mismatch direction from the zero mismatch case. The reason for this can be an intrinsic property of the synchronization case under study. Detailed study of this interesting question is beyond the scope of this paper.

It should also be underlined that in some cases there is a minor increase in the correlation coefficients with the increase of parameter mismatches. It could be due to the fluctuations, computer simulation results rounding error, etc.

Finally we dwell on the role of noise in synchronization. Noise can both improve and deteriorate the synchronization quality between the systems. It all depends on the intensity of noise and parameter mismatches between the systems. In the paper we study the influence of noise on the synchronization quality between the identical Josephson junctions. For this purpose we subject the right-hand side of Eqs. (1-12) to independent additive Gaussian white noise terms of form $\sigma\xi$, where $\sigma$ is the noise intensity and the correlation function for the noise terms is given by $\langle\xi(t)\xi(t_1)\rangle = 2\delta(t-t_1)$, where $\delta$ is the Dirac delta function.

Table 7 portrays the results of numerical simulation of Eqs.(1-12) with additive noise terms added to the right side of each equation and it demonstrates the dependence of the correlation coefficients on the noise intensity $\sigma$. Computing power used for the simulations presented here does not afford for the calculations of the correlation coefficients at higher noise intensities in the reasonable real-time simulation mode. It is clear from the data in Table 7 in the noise intensity range [0, $10^{-4}$] synchronization between the driven Josephson junctions is perfect; noise in this range does not affect the synchronization quality at all. As for the synchronization quality between the driver Josephson junction and driven Josephson junctions, the correlation coefficients at first are increasing, then reaching maximum at noise intensity $\sigma = 10^{-5}$ with subsequent decrease at higher noise intensities. It is noted that overall changes in the synchronization quality is around 1-2 %. Such a slight change occurs despite the changes of the noise intensity around the four orders.

**Table 1. Numerical simulation of Eqs. (1-12). Dependence of the correlation coefficients on the ratio $\tau_3\tau_1^{-1}$ and other parameters are as in Fig. 2 caption. Dimensionless units.**

| $\tau_3\tau_1^{-1}$ | $C_{1-2}$ | $C_{1-3}$ | $C_{1-4}$ | $C_{2-3}$ | $C_{2-4}$ | $C_{3-4}$ |
|---|---|---|---|---|---|---|
| 0.85 | 0.99 | 0.99 | 0.99 | 1 | 1 | 1 |
| 0.90 | 0.98 | 0.97 | 0.98 | 1 | 1 | 1 |
| 0.95 | 0.98 | 0.98 | 0.98 | 1 | 0.99 | 0.99 |
| 1 | 0.98 | 0.98 | 0.98 | 1 | 1 | 1 |
| 1.05 | 0.99 | 0.99 | 0.99 | 1 | 1 | 1 |
| 1.10 | 0.98 | 0.98 | 0.97 | 1 | 0.99 | 0.99 |
| 1.15 | 0.99 | 0.99 | 0.98 | 1 | 0.99 | 0.99 |

Table 2. Dependence of the correlation coefficients on the ratio $\alpha_{s3}\alpha_{s1}^{-1}$ and other parameters are as in Fig. 2 caption. Dimensionless units.

| $\alpha_{s3}\alpha_{s1}^{-1}$ | $C_{1-2}$ | $C_{1-3}$ | $C_{1-4}$ | $C_{2-3}$ | $C_{2-4}$ | $C_{3-4}$ |
|---|---|---|---|---|---|---|
| 0.85 | 0.99 | 0.99 | 0.99 | 1 | 0.99 | 0.99 |
| 0.90 | 0.99 | 0.99 | 0.98 | 1 | 0.99 | 0.99 |
| 0.95 | 0.99 | 0.99 | 0.99 | 1 | 0.99 | 0.99 |
| 1 | 0.98 | 0.98 | 0.98 | 1 | 1 | 1 |
| 1.05 | 0.99 | 0.99 | 0.99 | 1 | 0.99 | 0.99 |
| 1.10 | 0.98 | 0.98 | 0.99 | 1 | 0.99 | 0.99 |
| 1.15 | 0.98 | 0.98 | 0.99 | 1 | 0.99 | 0.99 |

Table 3. The correlation coefficients as a function of mismatch between $\beta_4$ and $\beta_2$ and other parameters are as in Fig. 2 caption. Dimensionless units.

| $\beta_4\beta_2^{-1}$ | $C_{1-2}$ | $C_{1-3}$ | $C_{1-4}$ | $C_{2-3}$ | $C_{2-4}$ | $C_{3-4}$ |
|---|---|---|---|---|---|---|
| 0.85 | 0.98 | 0.98 | 0.92 | 1 | 0.92 | 0.92 |
| 0.90 | 0.99 | 0.99 | 0.94 | 1 | 0.94 | 0.94 |
| 0.95 | 0.98 | 0.98 | 0.96 | 1 | 0.97 | 0.97 |
| 1 | 0.98 | 0.98 | 0.98 | 1 | 1 | 1 |
| 1.05 | 0.99 | 0.99 | 0.98 | 1 | 0.99 | 0.99 |
| 1.10 | 0.98 | 0.98 | 0.96 | 1 | 0.96 | 0.96 |
| 1.15 | 0.98 | 0.98 | 0.97 | 1 | 0.97 | 0.97 |

**Table 4. Dependence of the correlation coefficients on the mismatch between $i_{dc4}$ and $i_{dc2}$; other parameters are as in Fig. 2 caption. Dimensionless units.**

| $i_{dc4}i_{dc2}^{-1}$ | $C_{1-2}$ | $C_{1-3}$ | $C_{1-4}$ | $C_{2-3}$ | $C_{2-4}$ | $C_{3-4}$ |
|---|---|---|---|---|---|---|
| 0.85 | 0.99 | 0.99 | 0.97 | 1 | 0.97 | 0.97 |
| 0.90 | 0.98 | 0.98 | 0.96 | 1 | 0.96 | 0.96 |
| 0.95 | 0.98 | 0.98 | 0.97 | 1 | 0.98 | 0.98 |
| 1 | 0.98 | 0.98 | 0.98 | 1 | 1 | 1 |
| 1.05 | 0.99 | 0.99 | 0.98 | 1 | 0.98 | 0.98 |
| 1.10 | 0.98 | 0.98 | 0.96 | 1 | 0.96 | 0.96 |
| 1.15 | 0.99 | 0.99 | 0.93 | 1 | 0.93 | 0.93 |

**Table 5. The correlation coefficients as a function of the ratio $i_{04}i_{02}^{-1}$ and other parameters are as in Fig. 2 caption. Dimensionless units.**

| $i_{04}i_{02}^{-1}$ | $C_{1-2}$ | $C_{1-3}$ | $C_{1-4}$ | $C_{2-3}$ | $C_{2-4}$ | $C_{3-4}$ |
|---|---|---|---|---|---|---|
| 0.85 | 0.99 | 0.99 | 0.94 | 1 | 0.96 | 0.96 |
| 0.90 | 0.99 | 0.99 | 0.97 | 1 | 0.98 | 0.98 |
| 0.95 | 0.99 | 0.99 | 0.99 | 1 | 0.99 | 0.99 |
| 1 | 0.98 | 0.98 | 0.98 | 1 | 1 | 1 |
| 1.05 | 0.99 | 0.99 | 0.98 | 1 | 0.99 | 0.99 |
| 1.10 | 0.99 | 0.99 | 0.97 | 1 | 0.97 | 0.97 |
| 1.15 | 0.99 | 0.99 | 0.91 | 1 | 0.91 | 0.91 |

**Table 6. Dependence of the correlation coefficients on the ratio $\Omega_4\Omega_2^{-1}$ and other parameters are as in Fig.2 caption. Dimensionless units.**

| $\Omega_4\Omega_2^{-1}$ | $C_{1-2}$ | $C_{1-3}$ | $C_{1-4}$ | $C_{2-3}$ | $C_{2-4}$ | $C_{3-4}$ |
|---|---|---|---|---|---|---|
| 0.85 | 0.98 | 0.98 | 0.48 | 1 | 0.49 | 0.49 |
| 0.90 | 0.97 | 0.97 | 0.49 | 1 | 0.51 | 0.51 |
| 0.95 | 0.99 | 0.99 | 0.55 | 1 | 0.56 | 0.56 |
| 1 | 0.98 | 0.98 | 0.98 | 1 | 1 | 1 |
| 1.05 | 0.99 | 0.99 | 0.57 | 1 | 0.58 | 0.58 |
| 1.10 | 0.99 | 0.99 | 0.55 | 1 | 0.56 | 0.56 |
| 1.15 | 0.99 | 0.99 | 0.54 | 1 | 0.56 | 0.56 |

**Table 7. Numerical simulation of Eqs.(1-12) with additive noise terms added to the right side of each equation .Dependence of the correlation coefficients on the noise intensity. Parameters are as in Fig. 2 caption .Dimensionless units.**

| $\delta$ | $C_{1-2}$ | $C_{1-3}$ | $C_{1-4}$ | $C_{2-3}$ | $C_{2-4}$ | $C_{3-4}$ |
|---|---|---|---|---|---|---|
| 0 | 0.98 | 0.98 | 0.98 | 1 | 1 | 1 |
| $10^{-6}$ | 0.99 | 0.99 | 0.99 | 1 | 1 | 1 |
| $10^{-5}$ | 1 | 1 | 1 | 1 | 1 | 1 |
| $10^{-4}$ | 0.98 | 0.98 | 0.98 | 1 | 1 | 1 |

# 4 Discussion and Conclusions

As mentioned previously Terahertz radiation has immense potential in many scientific fields. As underlined in the Introductory Section most Terahertz generators require a lot of space, not cheap, not mobile and portable. Due to these factors the application areas of these generators could not be as wide as desired.

Here we dwell on the possibilities of the Terahertz waves for environmental applications [23]. We briefly highlight some examples of environmental applications of the Terahertz waves. As a first example we present the possibility of chemical sensing and substance identification by use of Teraherz imaging. It is well-know that the absorption of Terahertz frequencies is dominated by the excitation of intramolecular and intermolecular vibrations in substances. In other words, Terahertz vibrational spectra represent characteristic fingerprints for many chemical substances, as such can be used for contactless, non-invasive substance identification. Among them could be chemical warfare and biological warfare substances.

As a second example of Terahertz imaging in ecological applications we mention an explosive and land mine detection. From the ecological applications point of view of Terahertz waves one can also mention waste reduction possibility, say due the rust- under- paint detection. Briefly Terahertz imaging can be used to detect rust under paint without stripping off the paint. This is due to the fact that the rust causes surface roughness where the Terahertz radiation easily penetrate through the paint and detect the rough corroded surface. This application has important environmental benefit by eliminating the waste materials generated by stripping the paint and repainting large surface areas. This is particularly useful for automotive (cars, aircraft, and ships) manufacturers' service depots and building construction.

It is quite clear that all these ecological applications of the Terahertz waves require cost effective, compact, mobile, transportable Terahertz sources.

One of the most anticipated applications of the Terahertz imaging is in the area of medicine [9]. The thing is that the energy of the Tetahertz photon is around several or tens of milli eV; analogous energy for the X-ray photon is of the order of keV. In other words Terahertz radiation is not ionizing for the human being and allows for less hazardous imaging. At the same time one has to keep in mind that water is a good absorber for Terahertz waves. That is in – depth imaging could be a bit difficult. Still Terahertz imaging can be used for the detection of skin, mouth, throat and epithelelial cancers. Terahertz waves can also be used real time to confirm the removal of all cancer tissue, significantly reducing the need for subsequent operations. Terahertz imaging also allows for the intra-operative tool during breast cancer surgery. This allows the surgeon to check the removed tissue and to carry out further removals if required. Dentistry is another area in medicine where Terahertz imaging could allow for a good diagnostic via 3D imaging.

Again as in the case of ecological applications for the case of medicinal applications less spacious, affordable for the general public, mobile Terahertz sources are needed. Modeling conducted in [24] suggests that the micron-sized samples of high-temperature superconductors hold the potential of emitting at powers of ~1 mW under optimized conditions.

In this connection a search for inexpensive, cost effective mobile, compact Terahertz generators and detectors is a subject for intensive research endeavor [25-27]. Synchronization between the intrinsic numerous Josephson junctions in high-temperature superconductors could be a viable option from the above-mentioned point of view. In recent years the authors conducted an extensive research in the field of synchronization of Josephson junctions subject to the different coupling topology. The main result of this research was the fact that high quality synchronization between the Josephson junctions driven by a master junction is the optimal solution to achieve a high correlation between the driven junctions. In real life there are always some parameter differences between the junctions due to the fluctuations, technological processes, etc. From this point of view in real application perspectives the study of the effect of noise and parameter mismatches between the Josephson junctions on the quality of synchronization is of paramount importance.

Having in mind these perspectives in this research we have numerically studied the effect of parameter mismatches and noise on chaos synchronization quality between three uni-directionally time-delay coupled Josephson junctions. We have demonstrated the possibility of high quality synchronization between such systems. Although overall the parameter mismatches deteriorate the synchronization quality between the end junctions, in most sensitive cases high quality synchronization is still possible with 5-10 % mismatches. Deterioration of the synchronization quality under the influence of noise is also insignificant within the applied noise intensity range. The results are important for obtaining high power system of Josephson junctions, which is promising for the mismatch tolerant practical applications.

To summarize, at present as a rule available sources of terahertz radiation are spacious, not portable, not mobile, and not cost effective. Synchronization of Josephson junctions governed by a central junction could provide terahertz waves sources, which avoid these shortcomings. This property is extremely appealing from the application point of view.

## Acknowledgements

This work was supported by the Science Development Foundation under the President of the Republic Azerbaijan-Grant No EIF-KETPL-2-2015-1(25)-56/09/1.

## Competing Interests

Authors have declared that no competing interests exist.

___________________________________________________________________________